\documentclass[12pt]{reportj}
\usepackage{longtable}
\usepackage{listings}
\makeatletter
\def\@to{to}
\makeatother
\usepackage{times}
\usepackage{graphicx}
\usepackage{tocloft}
\usepackage{amssymb}
\usepackage{fancyheadings}
\usepackage{float}
\usepackage{xcolor} 
\lstdefinestyle{output}{backgroundcolor=\color{gray!10}}
\PassOptionsToPackage{hyphens}{url}
\usepackage{hyperref}
\usepackage[labelsep=space]{caption}

\emergencystretch=\maxdimen
\usepackage{datetime}
\newdateformat{ddmonthyyyy}{\THEDAY\ \monthname[\THEMONTH]\ \THEYEAR}

\renewcommand\thesection{\arabic{section}}
\renewcommand\thesubsection{\thesection.\arabic{subsection}}

\def\ssection#1{\setcounter{subsection}{0} \refstepcounter{section} \section*{\hbox to \hsize{\large\bf \arabic{section}. #1\hfill }}\label{sec} \addcontentsline{toc}{section}{\arabic{section}. #1}}
\def\ssubsection#1{\setcounter{subsubsection}{0} \refstepcounter{subsection}\subsection*{\hbox to \hsize{\normalsize\bfseries\itshape \arabic{section}.\arabic{subsection} #1\hfill}}\label{subsec} \addcontentsline{toc}{subsection}{\arabic{section}.\arabic{subsection} #1}}
\def\ssubsubsection#1{\refstepcounter{subsubsection}\subsection*{\hbox to \hsize{\normalsize\it \arabic{section}.\arabic{subsection}.\arabic{subsubsection} #1\hfill}}\label{subsubsec} \addcontentsline{toc}{subsubsection}{\arabic{section}.\arabic{subsection}.\arabic{subsubsection} #1}}

\def\ssectionstar#1{\section*{\hbox to \hsize{\large\bf #1\hfill}} \addcontentsline{toc}{section}{#1}}
\def\ssubsectionstar#1{\subsection*{\hbox to \hsize{\normalsize\bfseries\itshape #1\hfill}} \addcontentsline{toc}{subsection}{#1}}
\def\ssubsubsectionstar#1{\subsection*{\hbox to \hsize{\normalsize\it  #1\hfill}} \addcontentsline{toc}{subsection}{#1}}

\renewcommand{\cftaftertoctitle}{%
\mbox{}\hfill{\normalfont Page}}
\begin{document}

~\\

\vspace{-2.4cm}
\noindent\includegraphics*[width=0.295\linewidth]{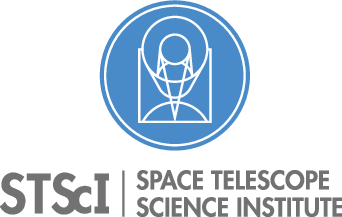}

\vspace{-0.4cm}

\begin{flushright}
    {\bf Instrument Science Report STIS 2026-07}
    
    \vspace{1.1cm}
    
    {\bf\Huge Identifying Cosmic Ray Over-Flagging in STIS CCD Spectroscopic Observations}
    
    \rule{0.25\linewidth}{0.5pt}
    
    \vspace{0.5cm}
    
    Matthew M. Dallas$^1$, Sean Lockwood$^1$
    \linebreak
    \newline
    \footnotesize{$^1$ Space Telescope Science Institute, Baltimore, MD}
    
    \vspace{0.5cm}
    
     30 September 2026
\end{flushright}

\vspace{0.1cm}

\noindent\rule{\linewidth}{1.0pt}
\noindent{\bf A{\footnotesize BSTRACT}}

The standard method for rejecting cosmic rays in CCD data taken by the Hubble Space Telescope's Space Telescope Imaging Spectrograph (STIS) instrument relies on combining multiple sub-exposures while flagging outliers. A known caveat of this method is that when the sub-exposures are not well registered, large portions of valid spectroscopic data can be flagged as cosmic rays. In this work we discuss the new \texttt{ocrreject\_exam()} task in the \texttt{stistools} Python package designed to quickly identify datasets impacted by this misregistration-induced over-flagging. We demonstrate its effectiveness on known cases of over-flagged data, introduce a new \href{https://www.stsci.edu/~STIS/monitors/cosmic_rays/}{monitoring website} which automatically runs the tool on STIS CCD spectroscopic data, and analyze the tool's output on STIS CCD spectroscopic data from 1997 through 2025. From this analysis we estimate that $\sim$31\% of datasets will exhibit cosmic ray over-flagging. Lastly, we discuss various methods of recalibrating impacted data to mitigate the effects of over-flagging.
{\it \noindent 
}

\vspace{-0.1cm}
\noindent\rule{\linewidth}{1.0pt}

\renewcommand{\cftaftertoctitle}{\thispagestyle{fancy}}
\tableofcontents


\vspace{-0.3cm}
\ssection{Introduction}\label{sec:Introduction}

Without the protection of much of the Earth's atmosphere, charge-coupled device (CCD) detectors on the Hubble Space Telescope are subject to a considerably larger number of space-borne energetic particles than CCD detectors in ground-based observatories. These cosmic rays have the effect of depositing charge on sections of pixels where the cosmic ray impacts the detector, creating areas of false counts not originating from the science target. The CCD detector on the Space Telescope Imaging Spectrograph (STIS) is no exception to these cosmic ray artifacts and, as such, part of the STIS calibration pipeline is designed to remove them from the processed data products. 

The algorithm used is described in detail in \href{https://www.stsci.edu/files/live/sites/www/files/home/hst/instrumentation/stis/documentation/instrument-science-reports/_documents/199618.pdf}{Baum et al. (1996)}, \href{https://www.stsci.edu/files/live/sites/www/files/home/hst/instrumentation/stis/documentation/instrument-science-reports/_documents/199811.pdf}{Shaw \& Hsu (1998)}, and \href{https://www.stsci.edu/files/live/sites/www/files/home/hst/instrumentation/stis/documentation/instrument-science-reports/_documents/199822.pdf}{Shaw \& Hodge (1998)}, but the general strategy is to split a single exposure into multiple sub-exposures or ``CR-Splits'' of roughly the same exposure time. These CR-Splits are corrected for large-scale bias then combined while contributions from outliers are rejected based on a noise model (effectively removing cosmic rays). The resulting combined image is then small-scale bias corrected, flat fielded, and dark subtracted, producing an image ready for spectral extraction. The pipeline products available for download from the \href{https://mast.stsci.edu/search/ui/#/hst}{Mikulski Archive for Space Telescopes} (MAST) for a given STIS visit or exposure include the raw data files, cosmic ray corrected/calibrated data, and the calibrated individual CR-Split exposures with the suffixes of \texttt{\_raw.fits}, \texttt{\_crj.fits}, and \texttt{\_flt.fits}, respectively. Both the \texttt{\_crj.fits} and \texttt{\_flt.fits} files include data quality (DQ) arrays which have been updated to set pixels flagged as cosmic rays with the flag value \href{https://hst-docs.stsci.edu/stisdhb/chapter-2-stis-data-structure/2-5-error-and-data-quality-array}{8192}.
This strategy for removing cosmic rays is effective when all CR-Splits are well registered with each other (within $\sim$0.1 CCD pixels in the spatial direction according to \href{https://www.stsci.edu/files/live/sites/www/files/home/hst/instrumentation/stis/documentation/instrument-science-reports/_documents/199822.pdf}{Shaw \& Hodge (1998)}). If this is not the case, the default rejection parameters can flag valid data from the observed spectrum in one or more of the CR-Splits as outliers and reject them as cosmic rays. When the cosmic ray rejected images are then co-added prior to the final spectral extraction, there is an artificial loss of flux where the real data has been ignored in some of the CR-Splits. 

In this work we present a new method in the \href{https://stistools.readthedocs.io/en/latest/}{\texttt{stistools}} Python package (currently the actively maintained code base for calibrating and manipulating STIS data) called \texttt{ocrreject\_exam()} to aid users in assessing whether a STIS CCD spectroscopic dataset has been affected by this cosmic ray over-flagging issue where real spectral data is flagged as cosmic rays in one or more CR-Splits. We also introduce a new \href{https://www.stsci.edu/~STIS/monitors/cosmic_rays/}{public monitoring website} which updates daily and runs the tool on all incoming STIS data. The monitor also displays the statistics and plots generated by the tool for all previous STIS CCD spectroscopic data for easy quick-looking of data. Lastly, we discuss potential steps to reprocess affected data to mitigate the impact of over-flagging. 


\lhead{}
\rhead{}
\cfoot{\rm {\hspace{-1.9cm} Instrument Science Report STIS 2026-07 Page \thepage}}

\vspace{-0.3cm}
\ssection{STIS Cosmic Ray Over-Flagging}\label{sec:cr_rejection}

As described in the Caveats section of \href{https://www.stsci.edu/files/live/sites/www/files/home/hst/instrumentation/stis/documentation/instrument-science-reports/_documents/199822.pdf}{Shaw \& Hodge (1998)}, CR-Splits taken with a spectroscopic grating that are misregistered by $\sim$0.1 pixels or more in the spatial direction are prone to having large portions of their spectral data flagged as cosmic rays in one or more of the exposure's CR-Splits. A general overview for why this occurs is that the cosmic ray rejection algorithm calculates an initial guess of the counts for a given pixel based on the median (or minimum, depending on the exposure time and number of CR-Splits) value of the pixel in the stack of CR-Splits. Pixels in a single CR-Split that are not within $\pm$ \texttt{sigma} $\cdot$ \texttt{noise} (where \texttt{sigma} is a parameter supplied in the CRREJTAB reference file and \texttt{noise} is a value which varies per pixel based on a \href{https://hst-docs.stsci.edu/stisdhb/chapter-3-stis-calibration/3-4-descriptions-of-calibration-steps#id-3.4DescriptionsofCalibrationSteps-3.4.53.4.5CRCORR:CosmicRayCorrection}{noise model}) are flagged as cosmic rays and their values are not used when determining the value of the pixel in the combined CR rejected image. In the case that more than one rejection cycle is specified, more iterations of cosmic ray flagging will occur with the guess and combined images updating based on the results of the previous cycle. The exact misregistration threshold for when CR over-flagging begins varies depending on the instrumental configuration and the properties of the target, but the principle remains the same: the algorithm assumes the bright pixels of the spectrum will not vary in the spatial direction between the CR-Splits. If the spectrum does shift, the counts in the pixels illuminated by the spectrum will vary more than the CR identification threshold based on the initial guess image and they will be flagged as cosmic rays in one or more splits. 

\begin{figure}[H]
  \centering
  \includegraphics[width=0.85\linewidth]{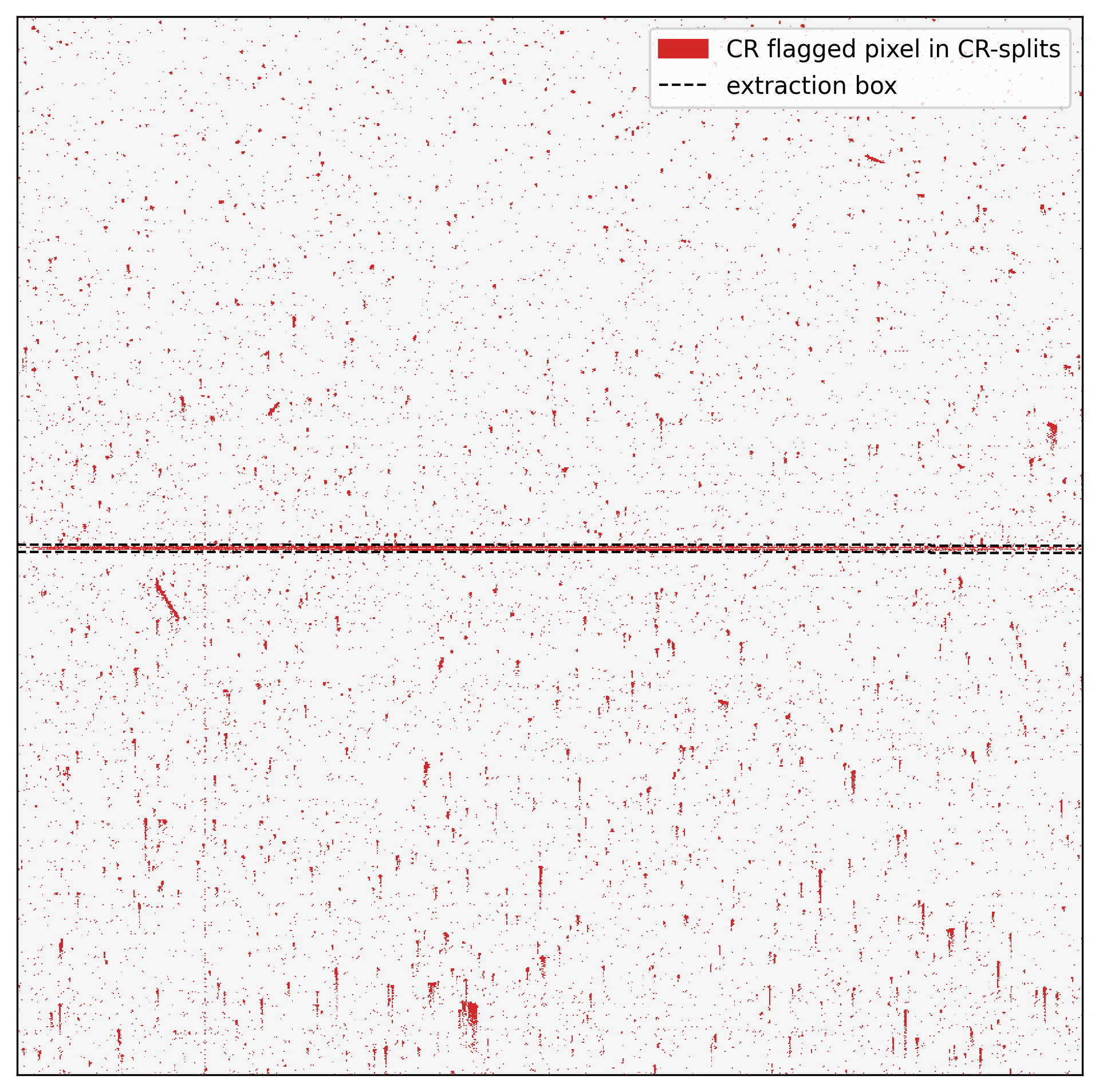}
    \caption{The CCD detector with pixels flagged as cosmic rays in either of the two CR-Splits in dataset \texttt{odvkl1030} colored in red. The extraction box located by \texttt{CALSTIS} is over-plotted with dashed black lines. The preferential flagging of pixels within the extraction box as cosmic rays was caused by the two CR-Splits being offset in this case by roughly 14 milliarcseconds in the spatial direction (or roughly 0.28 pixels at the CCD plate scale of 0.050872 arcseconds).} 
    \label{fig:crs_across_splits}
\end{figure}

As an example, in Figure~\ref{fig:crs_across_splits} every pixel flagged as a cosmic ray across the two CR-Splits that comprise the cosmic ray over-flagged dataset \texttt{odvkl1030} is plotted in red. The spectral extraction box is over-plotted with the dashed black line to give a sense of where the spectrum fell on the detector, and to demonstrate that a majority of the pixels inside it have been flagged as cosmic rays in at least one of the CR-Splits. 

\begin{figure}[H]
  \centering
  \includegraphics[width=\linewidth]{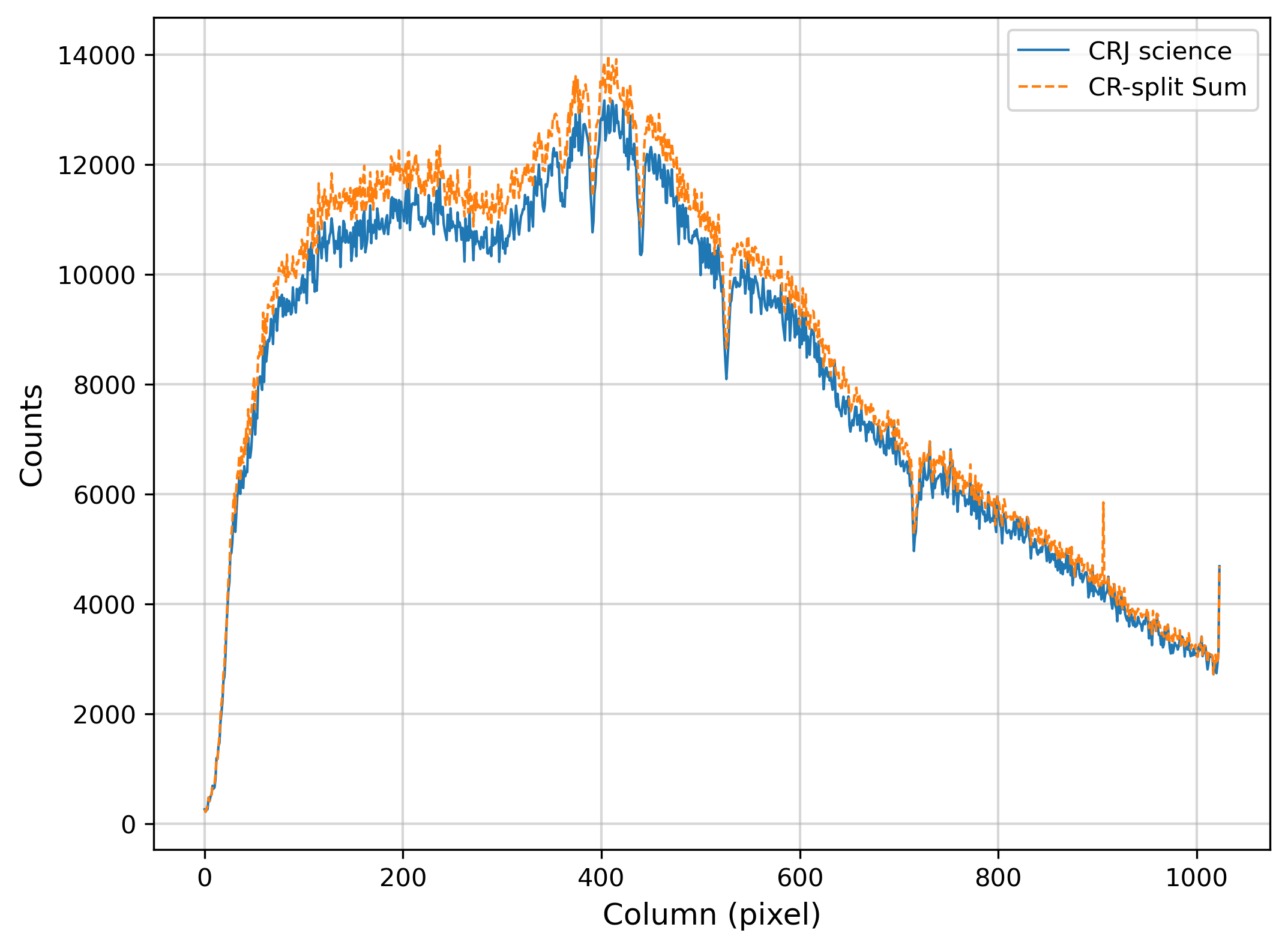}
    \caption{The counts recorded through the center row of the region used by \texttt{CALSTIS} to extract a spectrum (row 510 in this example) from the dataset \texttt{odvkl1030}. The counts of the row in the default \texttt{\_crj.fits} file are plotted in blue while the direct sum of the same row in both CR-Splits found in the \texttt{\_flt.fits} file is plotted in orange, showing how the cosmic ray over-flagging of bright pixels illuminated by the spectrum has caused the estimated counts to be biased lower. The end result of this is an artificial loss of flux in the extracted spectrum.}
    \label{fig:counts_comparison}
\end{figure}

When combined into a cosmic ray rejected image, the counts at these pixels will be estimated from what the algorithm determines to be an appropriately scaled value based on the non-rejected pixels in the stack. In the case of \texttt{odvkl1030}, this estimate is lower than would be the case if the algorithm had not flagged the spectrum in one or more CR-Splits as cosmic rays because the initial guess the cosmic ray rejection threshold is based on is set to be the minimum value of the stack. This ensures that the relatively brighter pixels are flagged and subsequently will not contribute to the applied scaling and resulting combined \texttt{\_crj.fits} file. This underestimation is demonstrated in Figure~\ref{fig:counts_comparison}, which plots the counts of the combined \texttt{\_crj.fits} file and the summed counts of both CR-Splits of the same dataset plotted in Figure~\ref{fig:crs_across_splits} through the center row of the extraction box. For more detailed discussions on the specifics of outlier rejection in the cosmic ray rejection algorithm of \texttt{CALSTIS}, see \href{https://www.stsci.edu/files/live/sites/www/files/home/hst/instrumentation/stis/documentation/instrument-science-reports/_documents/199618.pdf}{Baum et al. (1996)}, \href{https://www.stsci.edu/files/live/sites/www/files/home/hst/instrumentation/stis/documentation/instrument-science-reports/_documents/199811.pdf}{Shaw \& Hsu (1998)}, and \href{https://www.stsci.edu/files/live/sites/www/files/home/hst/instrumentation/stis/documentation/instrument-science-reports/_documents/199822.pdf}{Shaw \& Hodge (1998)}. 

\begin{figure}[H]
  \centering
  \includegraphics[width=\linewidth]{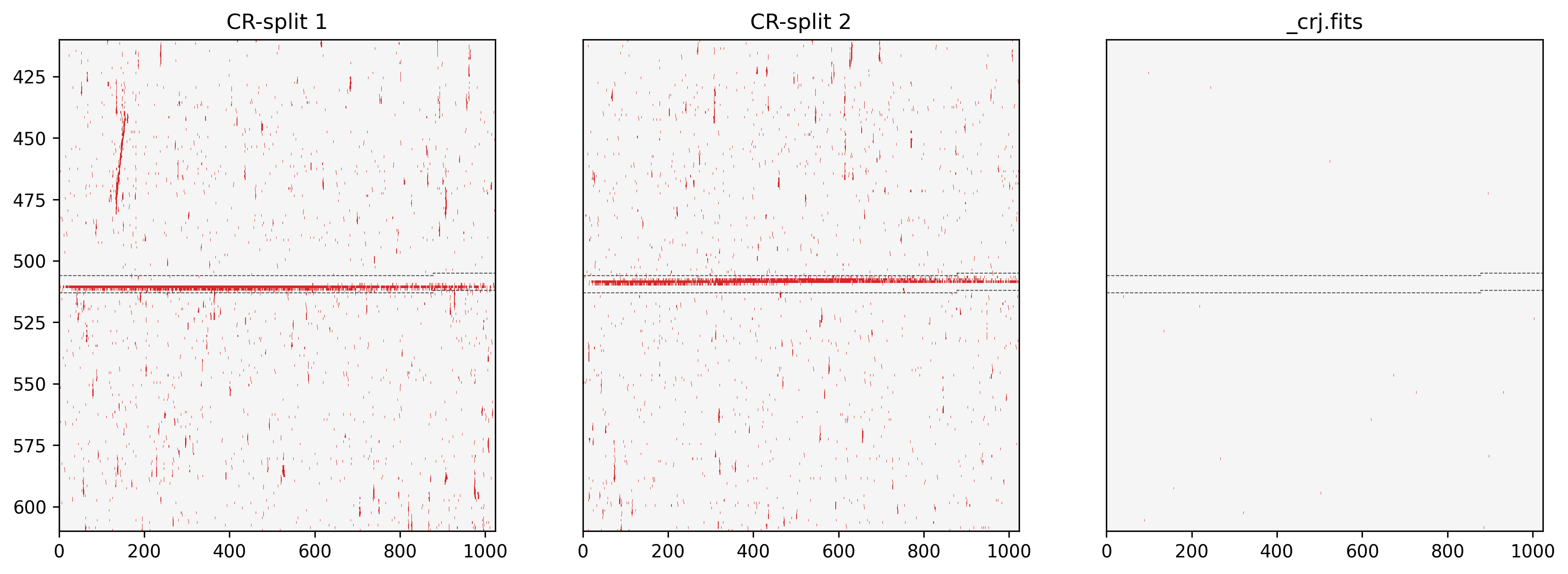}
    \caption{Plots of the CCD detector zoomed to $\pm$ 100 pixels above and below the center of the extraction box for CR-Split 1, CR-Split 2, and the resulting \texttt{\_crj.fits} file. Any pixel flagged as a cosmic ray (DQ = 8192) is plotted in red, while the extraction box region is over-plotted with a dashed line. In order for the 8192 flag to be set in the combined \texttt{\_crj.fits} file, the same pixel must be flagged with 8192 in all CR-Splits. This makes it difficult to determine based on the combined \texttt{\_crj.fits} file alone if a dataset was impacted by misregistration-induced cosmic ray over-flagging, and was the motivation for developing the \texttt{ocrreject\_exam()} tool to quickly identify such cases.}
    \label{fig:dq_hidden}
\end{figure}

What makes this problem elusive is that even if a pixel is flagged as a cosmic ray in one or more CR-Splits, the data quality array of the combined \texttt{\_crj.fits} only has the 8192 bit corresponding to ``data rejected in input pixel during image combination for cosmic ray rejection'' set for pixels that are flagged as cosmic rays across \textit{all} splits. It is therefore easy to look at the DQ extension of the \texttt{\_crj.fits} file and assume the algorithm did a good job of not mistakenly removing counts from a spectrum, when a large portion of the data in the extraction box has in fact been manipulated by the cosmic ray rejection algorithm under the assumption the target spectrum in one or more CR-Splits was cosmic rays. Figure~\ref{fig:dq_hidden} demonstrates this by showing the pixels identified as cosmic rays in the individual CR-Splits of the same dataset shown in Figure~\ref{fig:crs_across_splits}, alongside the pixels flagged as cosmic rays in the DQ array of the corresponding \texttt{\_crj.fits} file. A tool designed to identify datasets where this type of CR over-flagging occurs was first proposed in \href{https://www.stsci.edu/files/live/sites/www/files/home/hst/instrumentation/stis/documentation/instrument-science-reports/_documents/201902.pdf}{Carlberg (2019)} in response to the apparent loss of flux in the calibration visit associated with datasets \texttt{odvkl1020} and \texttt{odvkl1030}, and has now come to fruition in the form of the \texttt{ocrreject\_exam()} task and the \href{https://www.stsci.edu/~STIS/monitors/cosmic_rays}{STIS cosmic ray monitor}.

\vspace{-0.3cm}
\ssection{The \texttt{ocrreject\_exam()} Tool}\label{sec:ocrreject}

\ssubsection{Overview}\label{subsec:ocrreject_overview}

The \texttt{ocrreject\_exam()} tool is a Python routine designed to quickly identify STIS CCD spectroscopic datasets that have had cosmic rays over-flagged due to poor registration between CR-Splits. The routine operates on the principle that the fraction of pixels identified as cosmic rays inside and outside the default spectral extraction box throughout a series of CR-Splits should be roughly the same (i.e. there should not be an over-density of rejection among pixels the spectrum fell on). A preferential rejection of pixels within the extraction box of one or more CR-Splits indicates a significant failure of the cosmic ray rejection algorithm and is likely to have been caused by misregistration between CR-Splits as described in Section \ref{sec:cr_rejection}. The tool takes the associated \texttt{\_flt.fits} (flat-fielded data which contains the locations of the cosmic ray flagged pixels for each CR-Split) and \texttt{\_sx1.fits} (extracted spectrum which contains the location of the extraction box across the combined data) files for a dataset as inputs and outputs the fraction of pixels inside and outside the extraction box which were flagged as cosmic rays. This information is given for each CR-Split in addition to the average for the whole dataset. Particularly helpful for quickly identifying datasets impacted by CR over-flagging is the ratio of the average fraction of pixels flagged inside the extraction box to the average fraction of pixels flagged outside the box. As discussed further in Section \ref{sec:analysis}, if this \texttt{combined\_ratio} is above 1.0 the CR rejection algorithm has flagged more pixels inside the extraction box as cosmic rays than outside, which can be an indicator of misregistration-induced over-flagging. 

The first iteration of the \texttt{ocrreject\_exam()} tool was written as part of \href{https://www.stsci.edu/files/live/sites/www/files/home/hst/instrumentation/stis/documentation/instrument-science-reports/_documents/201902.pdf}{Carlberg (2019)} in response to apparent drops in the observed flux of a well-known standard white dwarf in datasets taken during proposal 15558 visit L1: part of the annual calibration program that monitors the sensitivity of the STIS CCD. For ease of comparison, the example usage of the new \texttt{ocrreject\_exam()} tool throughout this work will use the exposures taken during that same visit. Differences between this version of the code and the version presented in \href{https://www.stsci.edu/files/live/sites/www/files/home/hst/instrumentation/stis/documentation/instrument-science-reports/_documents/201902.pdf}{Carlberg (2019)} include:

\begin{itemize}
  \item Generalization to work with any CCD spectroscopic dataset (e.g. sub-array readouts, exposures which specified an \texttt{NRPTEXP} value instead of \texttt{CRSPLIT}, custom extracted files that generated an \_x1d.fits instead of an \_sx1.fits file, etc.)
  \item Static image and interactive plotting functionality which show the locations of cosmic ray flagged pixels across all CR-Splits on the whole detector, and the location of cosmic ray flagged pixels in each CR-Split around the extraction box.
  \item Output statistics for each individual CR-Split in addition to those of the whole exposure set. 
  \item Slight improvement to the calculation of the fraction of pixels flagged as cosmic rays outside the extraction box by directly referencing the DQ array rather than estimating a fraction across the whole detector via the \texttt{REJ\_RATE} and \texttt{TEXPTIME} header keywords.
  \item A statistical test for whether or not a dataset is over-flagged.
  \item Inclusion in the \texttt{stistools} Python package. 
\end{itemize}

\ssubsection{Installation}\label{subsec:installation}

The simplest way to install the \texttt{ocrreject\_exam()} tool is through the \texttt{stistools} Python package. The full version was incorporated into the package in release \texttt{1.4.9} in August 2026, so installing this or any later version of \texttt{stistools} will automatically install the \texttt{ocrreject\_exam()} tool as well. As described on the \texttt{stistools} \href{https://stistools.readthedocs.io/en/latest/gettingstarted.html}{Installation and Setup page}, the package is included in all releases of the Space Telescope Science Institute-maintained Conda environment: the Space Telescope Environment, or \texttt{stenv}. The \texttt{stenv} \href{https://stenv.readthedocs.io/getting_started.html}{Getting Started page} provides information on how to install the environment on a machine using any of the standard Micromamba, Mamba, or Conda environment managers. Setting up any \texttt{stenv} release $\geq$ \texttt{2026.09.18} will include a \texttt{stistools} release $\geq$ \texttt{1.4.9} and subsequently the \texttt{ocrreject\_exam()} tool.

The \texttt{stistools} package can also be installed directly into an existing virtual environment. Doing so would take the form of:

\begin{lstlisting}[language=bash]
(example_env) $ pip install "stistools>=1.4.9"
\end{lstlisting}

\noindent in a bash terminal with the virtual environment ``\texttt{example\_env}'' active.

As described in the \texttt{stistools} Installation and Setup, it is also possible to clone the package from the \href{https://github.com/spacetelescope/stistools.git}{GitHub page} to obtain the scripts locally, then install them into an environment. For further help on installing the \texttt{ocrreject\_exam()} tool please contact the \href{https://stsci.service-now.com/hst}{STIS helpdesk}.

\ssubsection{Syntax}\label{subsec:syntax}
Here we provide examples of the several ways the \texttt{ocrreject\_exam()} task can be run, meant to supplement the comprehensive descriptions available on the \href{https://stistools.readthedocs.io/en/latest/ocrreject_exam.html}{documentation page}. In Section \ref{sec:analysis} we discuss interpretations of the values and plots the \texttt{ocrreject\_exam()} tool generates.

Within Python, the task can be called as part of the \texttt{stistools} package. The most basic implementation would be to call the function within a directory containing the \texttt{\_flt.fits} and \texttt{\_sx1.fits} files for the dataset being inspected. (Note that the function \texttt{ocrreject\_exam()} is under the module \texttt{ocrreject\_exam}, consistent with other \texttt{stistools} routines). For example, the following Python 3.11 code would run the method on the dataset \texttt{odvkl1030}: 

\begin{lstlisting}[language=Python, basicstyle=\footnotesize\ttfamily]
>>> import stistools
>>> result = stistools.ocrreject_exam.ocrreject_exam('odvkl1030')
\end{lstlisting}

\noindent The output of \texttt{ocrreject\_exam()} is a list of dictionaries whose length is the number of input datasets. Printing the contents of the variable \texttt{result} from the above example therefore yields a list with a single dictionary containing the computed metrics for \texttt{odvkl1030}: 

\begin{lstlisting}[style=output]
[{'rootname': 'odvkl1030',
  'n_splits': 2,
  'detector_box_fraction': 0.0068359375,
  'n_cr_pix': array([12972, 12361]),
  'n_total_cr_pix': 25333,
  'extr_fracs': array([0.22544643, 0.25209263]),
  'outside_fracs': array([0.01090447, 0.01013436]),
  'combined_ratio': 22.697991784573254,
  'combined_ratio_threshold': 1.3895736431901895,
  'overflagged_stat': True}]
\end{lstlisting}

The full list of parameters is located on the \href{https://stistools.readthedocs.io/en/latest/ocrreject_exam.html}{documentation page} and is updated when changes to the module occur. Some particularly useful parameters are \texttt{data\_dir}, \texttt{plot}, and \texttt{verbose}. Setting these all to \texttt{True} would prompt the method to search in a specific directory for the relevant \texttt{\_flt.fits} and \texttt{\_sx1.fits} files, generate static diagnostic plots, and print short data summaries for each input dataset. For example:

\begin{lstlisting}[language=Python]
>>> from stistools.ocrreject_exam import ocrreject_exam

>>> datasets = ['ODVKL1020', 'ODVKL1030']
>>> directory = './visit_data'

>>> results = ocrreject_exam(obs_ids=datasets, data_dir=directory, plot=True, verbose=True)
\end{lstlisting}

\noindent would search the directory \texttt{./visit\_data} for the \texttt{\_flt.fits} and \texttt{\_sx1.fits} files associated with datasets \texttt{odvkl1020} and \texttt{odvkl1030}, populate a list of two Python dictionaries filled with cosmic ray over-flagging statistics using the identified files for each, store that list into the variable \texttt{results}, generate static \texttt{.png} diagnostic plots, and print a short description of each dataset to the console: 

\begin{lstlisting}[language=bash]
For ODVKL1020
Average rejection across all extraction boxes: 1.4%
Average rejection across all external regions: 0.9%
Combined ratio between the two: 1.45

For ODVKL1030
Average rejection across all extraction boxes: 23.9%
Average rejection across all external regions: 1.1%
Combined ratio between the two: 22.70
\end{lstlisting}

By default the \texttt{plot} argument stores the generated .png files into the directory specified by \texttt{data\_dir}, or the current working directory if \texttt{data\_dir} is not supplied. Interpretations of the outputs of the function are discussed in Section \ref{sec:analysis}.

The \texttt{ocrreject\_exam()} task can also be run via a command-line interface through the command \texttt{ocrreject\_exam} while within a virtual environment with the tool installed. For example, running:

\begin{lstlisting}[language=bash]
(example_env) $ ocrreject_exam odvkl1030
\end{lstlisting}

\noindent in a directory containing \texttt{odvkl1030\_flt.fits} and \texttt{odvkl1030\_sx1.fits} provides the following output in the command line:

\begin{lstlisting}[language=bash]
For odvkl1030
Average rejection across all extraction boxes: 23.9%
Average rejection across all external regions: 1.1%
Combined ratio between the two: 22.70
\end{lstlisting}

Calling the method in this way is essentially the same as running \texttt{ocrreject\_exam()} in the Python example with the \texttt{verbose} parameter set to \texttt{True}. All other parameters (plotting, data directory paths, etc.) are available via flags applied to the command. For a full list, see the \href{https://stistools.readthedocs.io/en/latest/ocrreject_exam.html}{documentation page} or run \texttt{ocrreject\_exam -h}.

\vspace{-0.3cm}
\ssection{The STIS CR Monitor Website}\label{sec:monitor}

Rather than requiring users to manually run the \texttt{ocrreject\_exam()} tool on data to perform a basic check on their data, we have set up a daily \href{https://www.stsci.edu/~STIS/monitors/cosmic_rays/}{monitoring website}\footnote{\url{https://www.stsci.edu/~STIS/monitors/cosmic_rays/}} which runs the \texttt{ocrreject\_exam()} tool on STIS CCD data when it is ingested into \href{https://mast.stsci.edu/search/ui/#/hst}{MAST}, as well as when new reference files are applied to older CCD data. The numerical and plot outputs of the tool are compiled into tables organized by year of observation for ease of use, along with additional information about each exposure such as CR-Split number, aperture, publicly available data quality flags, etc. The website holds tables including all of STIS's past CCD observations going back to the start of observing in 1997. Each table is linked via the home page, which also displays descriptions of each column present in the yearly tables under the ``Column Definitions'' section. Further discussion of the values displayed in the yearly tables is presented in Section \ref{sec:analysis}. Note that although this monitor is public facing, the data displayed provides no spectroscopic information.

\vspace{-0.3cm}
\ssection{Analysis of Cosmic Ray Over-Flagging Statistics}\label{sec:analysis}

\ssubsection{Identification of Over-Flagged Datasets}\label{subsec:overflagged_stat}

So far we have discussed how cosmic ray over-flagging occurs via the misregistration of CR-Splits, how this over-flagging can lead to incorrect count estimates within the combined data, and have introduced the \texttt{ocrreject\_exam()} task and the publicly available \href{https://www.stsci.edu/~STIS/monitors/cosmic_rays/}{STIS CCD Cosmic Ray Monitor} website as new tools to identify and analyze cosmic ray over-flagged STIS data. This section will discuss the specifics of the method the tool uses to identify datasets with cosmic ray over-flagging. 

The \texttt{overflagged\_stat} output of \texttt{ocrreject\_exam()} reports whether the fraction of cosmic ray flagged pixels observed in the extraction box could be caused by a truly random distribution of cosmic rays across the detector within a provided confidence level. The test uses \texttt{scipy}'s \texttt{binomial} module to compute the maximum \texttt{combined\_ratio} result expected under the null hypothesis that cosmic rays are distributed randomly across the detector. This value is reported as the \texttt{combined\_ratio\_threshold}. 
A dataset with an observed \texttt{combined\_ratio} is labeled as over-flagged and highlighted in orange on the monitor website if this \texttt{combined\_ratio\_threshold} is less than or equal to the observed \texttt{combined\_ratio}. The number of pixels each real cosmic ray is expected to impact can be controlled via the \texttt{cr\_size} parameter (increasing this value effectively increases the believable \texttt{combined\_ratio\_threshold} because there is a higher probability of observing more cosmic ray flagged pixels in the extraction box if more than one pixel can be accounted for by a single hit). The confidence level used in the calculation can also be controlled via the \texttt{alpha} parameter.
By default these parameters are set to 2.4 pixels per cosmic ray and 0.00135, which are the mean number of pixels flagged as cosmic rays per cluster in the datasets analyzed in Section Section~\ref{subsec:performance_on_data}, and the upper tail of a two-sided 99.73\% confidence interval (equivalent to the upper tail of a Gaussian 3-sigma) respectively. These defaults are empirically verified in Section \ref{subsec:performance_on_data}.

\ssubsection{Performance on Real Data}\label{subsec:performance_on_data}

\begin{figure}[H]
  \centering
  \includegraphics[width=0.75\linewidth]{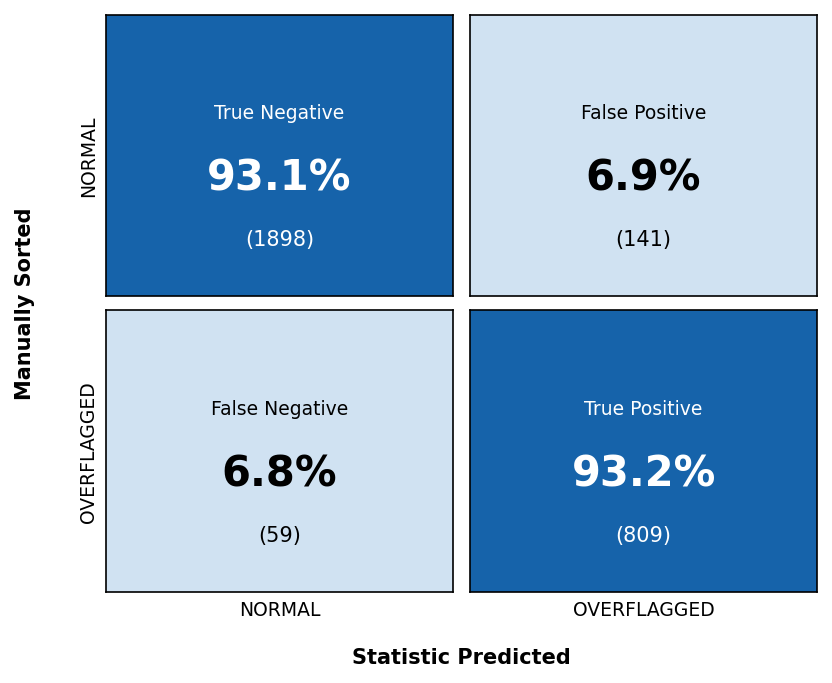}
    \caption{The confusion matrix resulting from comparing the \texttt{overflagged\_stat} prediction to the manual classification of the 2,907 CCD spectra. Note that the matrix is row-normalized where the percentages shown are of the manually sorted ``true'' values, i.e. there are 809+59=868 over-flagged datasets, 59 of which are considered normal based on the automated \texttt{overflagged\_stat} prediction.}
    \label{fig:confusion_matrix}
\end{figure}

To quantify how well this method identifies over-flagged data we manually sorted the STIS CCD spectra on the monitor website from the beginning of 2016 through the end of 2025 into ``normal'' and ``over-flagged'' samples using the monitoring plots and checking for the characteristic flagged spectral cores like that shown in Figure~\ref{fig:crs_across_splits}. This sample consisted of 2,907 observations, 2,039 of which we categorized as ``normal'' and 868 as ``over-flagged''. We find our manual labels agree with the prediction made from \texttt{ocrreject\_exam()}'s \texttt{overflagged\_stat} on 2,707 out of the 2,907 datasets ($\sim$93\%) with our chosen default values. We show the resulting confusion matrix in Figure~\ref{fig:confusion_matrix}.

Although we find that marginally higher overall accuracies can be achieved ($\sim$96\%) when setting the \texttt{cr\_size} parameter higher, to 6 pixels per cosmic ray as is discussed in Section 4 of \href{https://www.stsci.edu/files/live/sites/www/files/home/hst/instrumentation/stis/documentation/instrument-science-reports/_documents/199822.pdf}{Shaw \& Hodge (1998)}, this is at the cost of increasing the relative amount of false negatives (datasets incorrectly predicted to be ``normal''). In a monitoring context, we choose to allow more ``normal'' datasets to be incorrectly labeled in order to not miss ``over-flagged'' data in the monitor. We note that the false negatives occur more frequently in datasets with more total cosmic rays/longer exposure times ($\gtrsim$20,000 total cosmic ray flagged pixels and $\gtrsim$200s total exposure time respectively), while the false positives (datasets incorrectly predicted to be ``over-flagged'') occur more frequently in datasets with fewer total cosmic rays/shorter exposure times. We also find that both the false positives and false negatives occur when the observed \texttt{combined\_ratio} is within $\pm$0.75 of the \texttt{combined\_ratio\_threshold}, while true positives often occur when the \texttt{combined\_ratio} is $\gg$ than the \texttt{combined\_ratio\_threshold}. Other physical reasons why the metric can fail include false negatives when the over-flagged spectral core is outside of the default extraction box, and false positives when legitimate grazing cosmic rays cross through the extraction box. It is therefore advisable to check the diagnostic plots in addition to the \texttt{overflagged\_stat} to determine the disposition of a given dataset.

\begin{figure}[H]
  \centering
  \includegraphics[width=\linewidth]{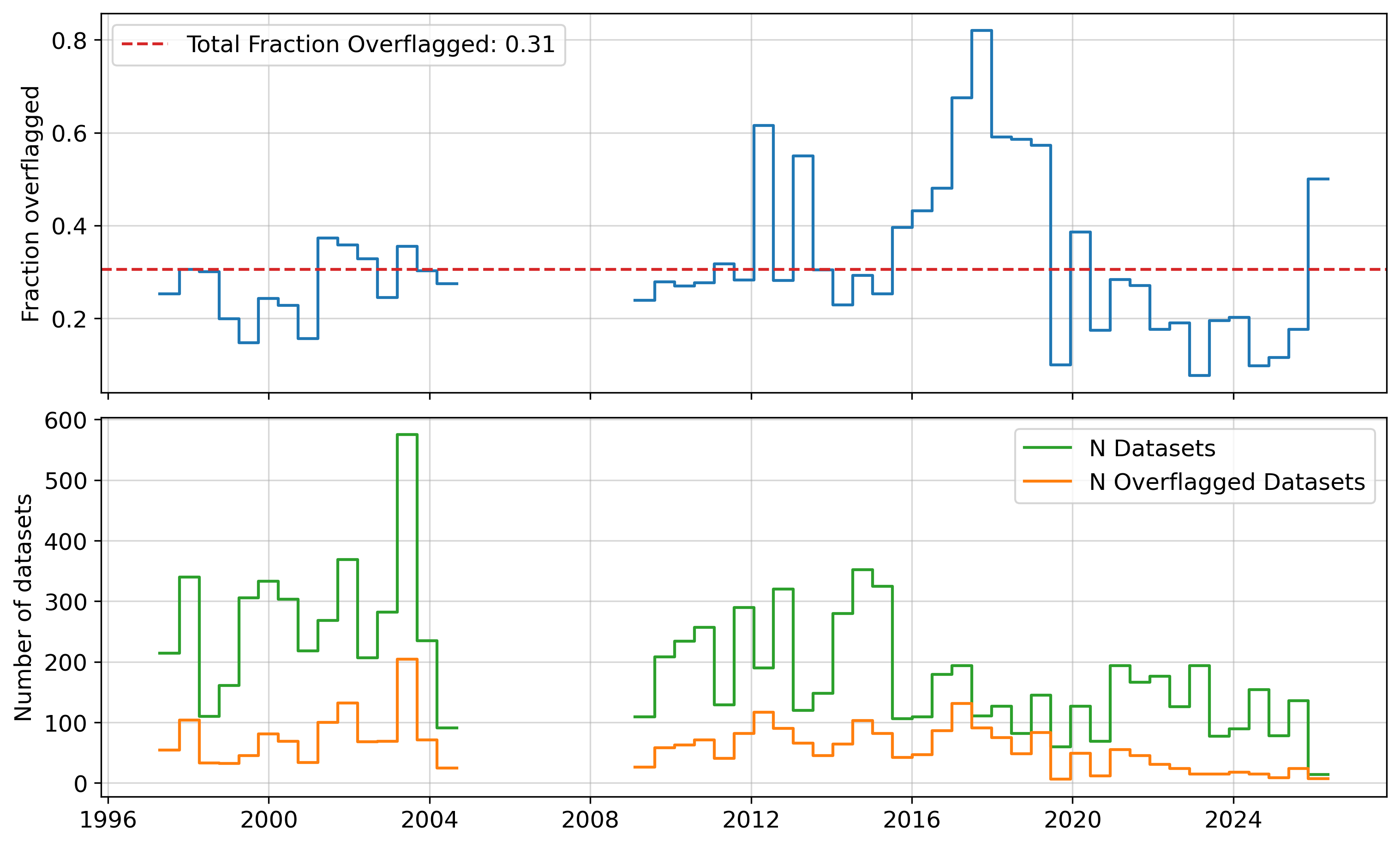}
    \caption{Top plot: the fraction of datasets with \texttt{overflagged\_stat} = \texttt{True} in 180-day bins since the first STIS CCD Spectroscopic dataset. Bottom plot: the number of datasets in each 180-day bin. For these samples we filtered down to only those datasets with no bad quality comments to avoid other observatory failures such as losses of lock.}
    \label{fig:fraction_over_time}
\end{figure}

Based on the predictions made by the \texttt{overflagged\_stat} on all the datasets run by the monitor from 1997 through 2025 associated with no bad quality comments, we estimate that $\sim$31\% of spectroscopic CCD data are cosmic ray over-flagged (i.e. would be classified in the right column of Figure~\ref{fig:confusion_matrix}). This is consistent with the discussion in \href{https://www.stsci.edu/files/live/sites/www/files/home/hst/instrumentation/stis/documentation/instrument-science-reports/_documents/199618.pdf}{Baum et al. (1996)}, where the designers of the cosmic ray rejection algorithm anticipated that $\gtrsim$20\% of users would want to re-perform the cosmic ray rejection step to improve upon the pipeline. However, as demonstrated in Figure~\ref{fig:fraction_over_time} this value is time dependent. For example, the period from around 2016 to 2019 with higher rates of cosmic ray over-flagging is correlated with the period of higher pointing instability discussed in the \href{https://www.stsci.edu/contents/news/stis-stans/july-2018-stan.html#h3-CK-aa3b1980-82c6-4009-86c0-4216e8db8b76}{July 2018 STIS newsletter}.  

As shown in Section 3.2 of \href{https://www.stsci.edu/files/live/sites/www/files/home/hst/instrumentation/stis/documentation/instrument-science-reports/_documents/201902.pdf}{Carlberg (2019)}, the 2-D histograms of the pointing in the \texttt{\_jif.fits} files can also be used to identify exposures where the pointing is likely to move the core of a spectrum on the detector. In the future another quantitative method using these files such as checking the FWHM of the pointing distributions could be implemented alongside the \texttt{overflagged\_stat} as an additional check for datasets with misregistration-induced CR over-flagging.

\vspace{-0.3cm}
\ssection{Correcting for Cosmic Ray Over-Flagging}\label{sec:correcting_data}

\ssubsection{Basic Recalibration}\label{subsec:parameters}
Fortunately, many cases of misregistration-induced cosmic ray over-flagging can be mitigated by reprocessing the data with adjustments to the default parameter values of the cosmic ray algorithm. This can be accomplished using the \texttt{stistools} routine \texttt{ocrreject}, which is designed to add STIS exposures while rejecting cosmic rays. The \href{https://stistools.readthedocs.io/en/latest/ocrreject.html}{documentation page} has a full description of the capabilities of \texttt{stistools.ocrreject()}. However, because the effect each parameter has on an individual dataset varies depending on the observation, here we take the over-flagged dataset \texttt{odvkl1030} and use \texttt{stistools.ocrreject()} to apply a few very basic mitigation strategies and show that the amount of preferentially flagged cosmic rays inside the extraction box decreases with each. 

The default cosmic ray rejection parameters listed in the primary header of the \_crj.fits file of \texttt{odvkl1030} are:

\begin{lstlisting}[style=output, basicstyle=\scriptsize\ttfamily, breaklines=false]
/ COSMIC RAY REJECTION ALGORITHM PARAMETERS                          
MEANEXP   = 8.6400002E+01   / reference exposure time for parameters           
SCALENSE  = 0.0000000E+00   / multiplicative scale factor applied to noise     
INITGUES  = 'minimum '      / initial guess method (MIN or MED)                
SKYSUB    = 'none    '      / sky value subtracted (MODE or NONE)              
CRSIGMAS  = '4.0     '      / statistical rejection criteria                   
CRRADIUS  = 1.5000000E+00   / rejection propagation radius (pixels)            
CRTHRESH  = 7.5000000E-01   / rejection propagation threshold                  
BADINPDQ  = 0               / data quality flag bits to reject                 
REJ_RATE  = 1.1296875E+02   / rate at which pixels are affected by cosmic ray  
CRMASK    = T               / flag CR-rejected pixels in input files (T/F)     
\end{lstlisting}

\begin{figure}[H]
  \centering
  \includegraphics[width=\linewidth]{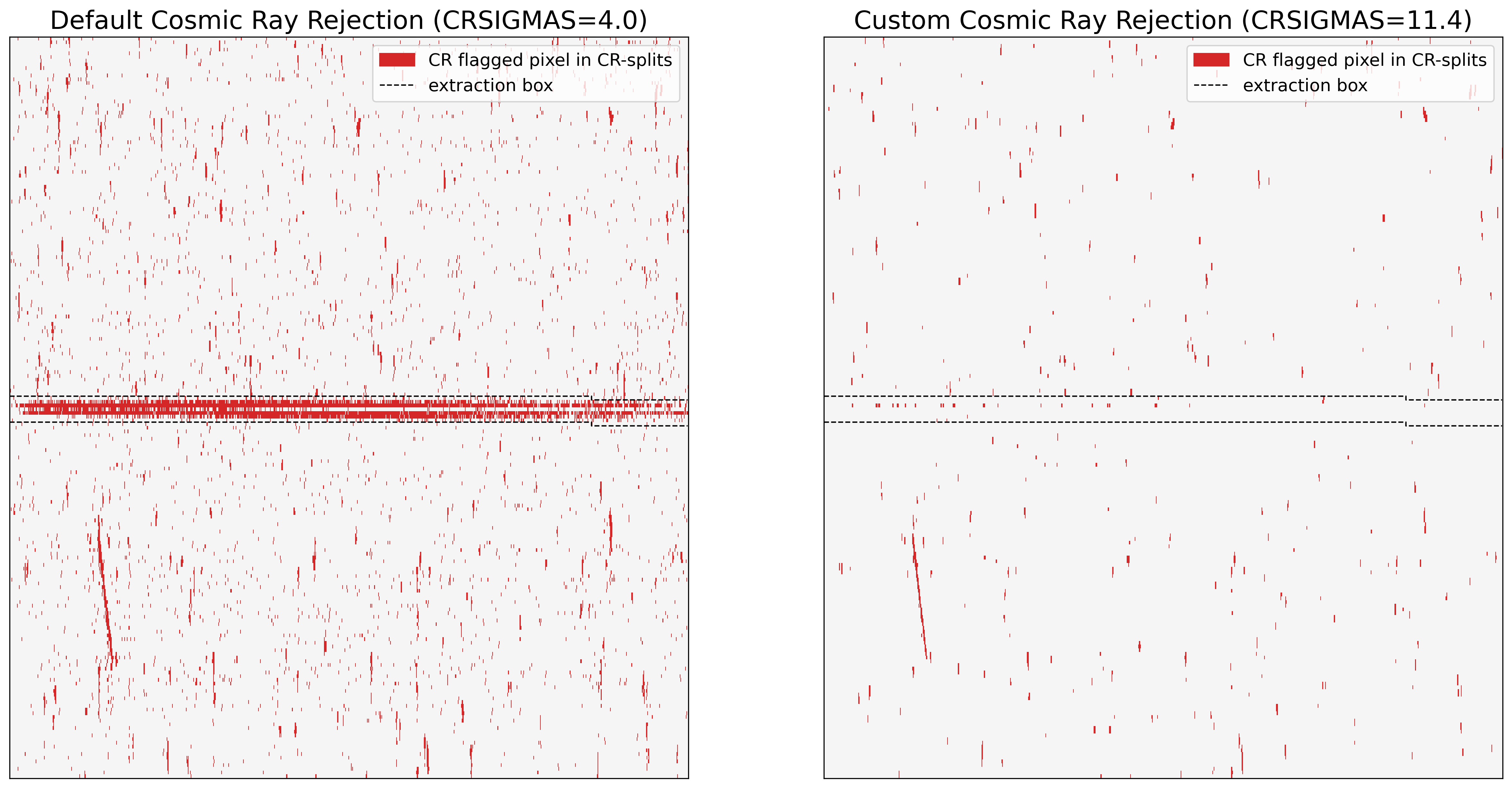}
    \caption{The CCD detector (zoomed to the rows $\pm$ 100 pixels of the extraction center) with pixels flagged as cosmic rays in either of the two CR-Splits in dataset \texttt{odvkl1030} before and after implementing a higher \texttt{CRSIGMAS} value in \texttt{stistools.ocrreject()}. Increasing the \texttt{CRSIGMAS} value allows for most of the spectrum to correctly not be counted as cosmic rays. Note that increasing the \texttt{CRSIGMAS} value can allow real cosmic rays to not be flagged and thus get included in the final combined image. This may be deemed acceptable to recover lost flux, as is the case in \texttt{odvkl1030}.}
    \label{fig:increase_sigma}
\end{figure}

Running \texttt{ocrreject\_exam} on the default products for \texttt{odvkl1030} returns a \texttt{combined\_ratio} value of $\sim$23, and as shown in Figures~\ref{fig:crs_across_splits} and ~\ref{fig:counts_comparison} the pixels in the extraction box across both exposures are practically all flagged as cosmic rays and the combined counts estimate is diminished as a result. Increasing the \texttt{CRSIGMAS} value (in this case substantially due to the brightness of the source) from 4.0 to 11.4 returns a \texttt{combined\_ratio} value of $\sim$1.4 when the data products are re-run through \texttt{ocrreject\_exam}. A comparison of the pixels flagged in either CR-Split between the pipeline-produced files and the reprocessed files is shown in Figure~\ref{fig:increase_sigma}.

Similarly, changing the \texttt{INITGUES} parameter from ``minimum'' to ``median'' alone reduces the \texttt{combined\_ratio} value to $\sim$15. Lowering both the \texttt{CRRADIUS} (how many pixels around an identified cosmic ray pixel to also consider excluding on a lower threshold) and \texttt{CRTHRESH} (the scale this lower neighboring exclusion threshold is set at) can also mitigate cosmic ray over-flagging inside the extraction box, as well as introducing rejection cycles (see \href{https://www.stsci.edu/files/live/sites/www/files/home/hst/instrumentation/stis/documentation/instrument-science-reports/_documents/199822.pdf}{\S4 of Shaw \& Hodge (1998)}). For \texttt{odvkl1030}, either lowering the \texttt{CRRADIUS} to a value between 0 and 1 or changing the \texttt{CRTHRESH} to 1.0 to cause the cosmic ray rejection algorithm to either not consider nearby pixels for flagging at a lower threshold or make that threshold the same as when the pixels are normally flagged respectively, reduces the \texttt{combined\_ratio} value to $\sim$16.4.

The parameters in \texttt{stistools.ocrreject()} are not mutually exclusive. For example, increasing the \texttt{CRSIGMAS} value to 5.6 while changing the \texttt{INITGUES} parameter to ``median'' also yields a \texttt{combined\_ratio} value of $\sim$1.3. Again, the parameters will have different effects on different datasets depending on the nature of the exposures, so it is best to tailor the parameters around a given dataset rather than adopt a single mitigation strategy. This is especially true given the results also change when the data are processed starting with the \texttt{\_flt.fits} file compared to the \texttt{\_raw.fits} file, as is discussed in Section~\ref{subsec:crreject_from_raw}.

\ssubsection{The \texttt{crrej\_from\_raw()} Tool}\label{subsec:crreject_from_raw}

The \texttt{stistools.ocrreject()} tool described above reads in bias-corrected, dark-subtracted, and flat-fielded \texttt{\_flt.fits} files to perform CR-combination, which differs from the standard order-of-processing applied by \texttt{CALSTIS}, and thus used in pipeline products. \texttt{CALSTIS} performs CR-rejection after the \texttt{BIASCORR} step, as seen in \href{https://hst-docs.stsci.edu/stisdhb/chapter-3-stis-calibration/3-3-data-flow-through-calstis#id-3.3DataFlowThroughcalstis-fig3.5}{\S3.3 of the STIS Data Handbook} (Rickman, E., Brown, J., et al., 2024). To facilitate performing modified CR-rejection in the same order, we have recently released a tool, \href{https://stistools.readthedocs.io/en/latest/crrej_from_raw.html}{\texttt{stistools.crrej\_from\_raw()}}, that allows users control over the same parameters described above by creating and assigning a temporary \texttt{CCDTAB} reference file and processing starting with raw data products.

According to Shaw \& Hodge (1998), \texttt{CALSTIS} performs CR-rejection before flat-fielding to avoid poorer CR-rejection from large-scale spatial variations (e.g. vignetting) and to properly propagate errors from the individual exposures. While the two CR-rejection methods produce similar results, users are advised to use \texttt{crrej\_from\_raw} when the most accurate results are required (see both \href{https://www.stsci.edu/files/live/sites/www/files/home/hst/instrumentation/stis/documentation/instrument-science-reports/_documents/199822.pdf}{\S3 of Shaw \& Hodge (1998)} and \href{https://www.stsci.edu/files/live/sites/www/files/home/hst/instrumentation/stis/documentation/instrument-science-reports/_documents/199618.pdf}{\S4 of Baum et al. (1996)}).

\vspace{-0.3cm}
\ssection{Conclusion}\label{sec:conclusion}
 
Despite the robustness of the STIS cosmic ray rejection algorithm, misregistration of CR-Splits can cause a substantial amount of valid data to be flagged as cosmic rays. Although this problem is difficult to catch through inspecting \texttt{\_crj.fits} files alone, the new \texttt{ocrreject\_exam()} tool can now be used to quickly identify when a dataset has pixels preferentially flagged as cosmic rays inside the extraction box by using both diagnostic plots and a built-in statistical test. This task is available in \texttt{stistools} versions $\geq$ \texttt{1.4.9} and  \texttt{stenv} releases $\geq$ \texttt{2026.09.18}. The outputs of the tool with default settings are also displayed on the publicly available STIS CCD Cosmic Ray Monitor \href{https://www.stsci.edu/~STIS/monitors/cosmic_rays/}{website}. Based on all STIS CCD spectroscopic data available from 1997-2025 on this monitor, we find that around 31$\%$ of STIS CCD spectroscopic datasets are over-flagged. In many cases of misregistration-induced over-flagging, recalibrating the data with the \texttt{stistools.ocrreject()} or \texttt{stistools.crrej\_from\_raw()} tools can help reduce the number of incorrectly flagged pixels.


\vspace{-0.3cm}
\ssectionstar{Acknowledgements}
\vspace{-0.3cm}
We thank Joleen Carlberg, Serge Dieterich, and Aidan Pidgeon for their expertise and advice regarding the STIS cosmic ray rejection algorithm and data analysis and their thoughtful reviews.

\vspace{-0.3cm}
\ssectionstar{References}\label{sec:References}
\vspace{-0.3cm}

\noindent
Carlberg, J., 2019, STIS Instrument Science Report 2019-02\\
Dick Shaw, \& Phil Hodge, 1998, STIS Instrument Science Report 1998-22\\
R. Shaw, \& J. C. Hsu, 1998, STIS Instrument Science Report 1998-11\\
Rickman, E., Brown J., et al., 2024, STIS Data Handbook, Version 8.0, (Baltimore: STScI).\\
Stefi Baum et al. 1996, STIS Instrument Science Report 1996-18\\

\end{document}